\documentclass[aps,pre,twocolumn,amsmath,amssymb]{revtex4-1}

\usepackage{graphicx,txfonts}
\usepackage{multirow,dcolumn}

\begin{document}

\title{Navigating temporal networks}

\author{Sang Hoon Lee}
\affiliation{Department of Liberal Arts, Gyeongnam National University of Science and Technology, Jinju 52725, Korea}
\affiliation{School of Physics, Korea Institute for Advanced Study, Seoul 02455, Korea}

\author{Petter Holme}
\email{holme@cns.pi.titech.ac.jp}
\affiliation{Institute of Innovative Research, Tokyo Institute of Technology, Nagatsuta-cho 4259, Midori-ku, Yokohama, Kanagawa, 226-8503, Japan}

\begin{abstract}
Navigation on graphs is the problem how an agent walking on the graph can get from a source to a target with limited information about the graph. The information and the way to exploit it can vary. In this paper, we study navigation on temporal networks---networks where we have explicit information about the time of the interaction, not only who interacts with whom. We contrast a type of greedy navigation---where agents follow paths that would have worked well in the past---with two strategies that do not exploit the additional information. We test these on empirical temporal network data sets. The greedy navigation is indeed more efficient than the reference strategies, meaning that there are correlations in the real temporal networks that can be exploited. We find that the navigability for individual nodes is most strongly correlated with degree and burstiness, i.e., both topological and temporal structures affect the navigation efficiency.
\end{abstract}

\maketitle

\section{Introduction}
\label{sec:introduction}

Temporal networks can be seen as an extension of the static network paradigm to include information about when interactions happen, not only between whom~\cite{TemporalNetworkReview1,TemporalNetworkReview2,masudalambiotte}. Just as for static networks, it is interesting to study dynamic phenomena happening on temporal networks, and how the structure of the interaction affects these. In the literature, there has been a great focus on disease spreading~\cite{masuda_holme_rev}. Somewhat less commonly, researchers have studied random walks~\cite{Delvenne2010,Perra2012,Starnini2012,Hoffmann2012,Ribeiro2013,Lambiotte2013,Rocha2014,Mata2014,Speidel2015,Delvenne2015} and threshold models of social spreading phenomena~\cite{taka,PhysRevE.89.062815,KARIMI20133476}. Another fundamental dynamic problem on networks is navigation~\cite{Kleinberg2000,SHLee2012}. This concerns agents traveling on the network with given starting points and destinations, but with incomplete knowledge of the network, like the sense of direction in spatially embedded networks~\cite{SHLee2012}. The problem of navigation on temporal networks has so far not been explored in the literature. The goal of this paper is to establish this research question and investigate solutions in form of an extension of spatially navigating agents.

The basic setting is a stream of contacts---triples $(i,j,t)$ of two nodes $i$ and $j$ and a time $t$---representing an interaction event between the two nodes. Then we assume an agent, as a walker in a random walk, can move to another node at the time of a contact. When a contact happens, we assume that the agent can make the decision whether or not to take a step, based on the history of the temporal network. In line with the assumption of incomplete information, we assume that future contacts are not known to a node. For simplicity, however, we assume the last observed time from destination to target is obtainable for all nodes. In this setting, we test three strategies. One is called \textit{greedy navigation} (GN) where agents jump from $i$ to $j$ at a contact $(i,j,t)$ if the previously observed time to reach the target is shorter from $j$ than $i$. The other two strategies are for reference and not using any available information. The second strategy is the \textit{greedy walk} (GW) of Ref.~\cite{Saramaki2015} where agents move at every contact. The third strategy is to simply wait at the origin until there is a contact with the target, we call it the \textit{wait for target} (WFT) strategy.

We explore the strategies on six empirical temporal networks. The reason why we use empirical contact data rather than temporal-network models as the basis of our work is twofold. First, there is a very large number of possible structures and correlations in temporal networks compared with static networks, so that one cannot simply scan through them in models~\cite{TemporalNetworkReview2}. It is also very challenging to identify the most important structures for the dynamic process in question~\cite{TemporalNetworkReview2}. Second, studying empirical networks contributes to the understanding of the original system itself. Such an analysis enables us how different data sets differ with respects to navigating agents.

In  the remainder of this paper, we will present and motivate the model of navigation, present the empirical data sets, our analysis procedure, and our simulation results.

\section{Preliminaries}
\label{sec:model}

\subsection{Temporal network representations}

There are different ways to incorporate information about the timing of contacts into network modeling. In this work, we will use so-called the \textit{contact list} (sometimes the \textit{link stream}) framework~\cite{TemporalNetworkReview1}. In that setting, the basic unit of interaction is a \textit{contact} (or \textit{event})---a triple $(i,j,t)$ showing that nodes $i$ and $j$ are in contact at time $t$. The time from the first to last time in a data set is the \textit{duration} $T$. Other descriptive quantities are the time resolution (minimal time between two contacts) $\delta t$, the number of contacts $C$, and the number of nodes $N$.

\begin{figure}
\includegraphics[width=\columnwidth]{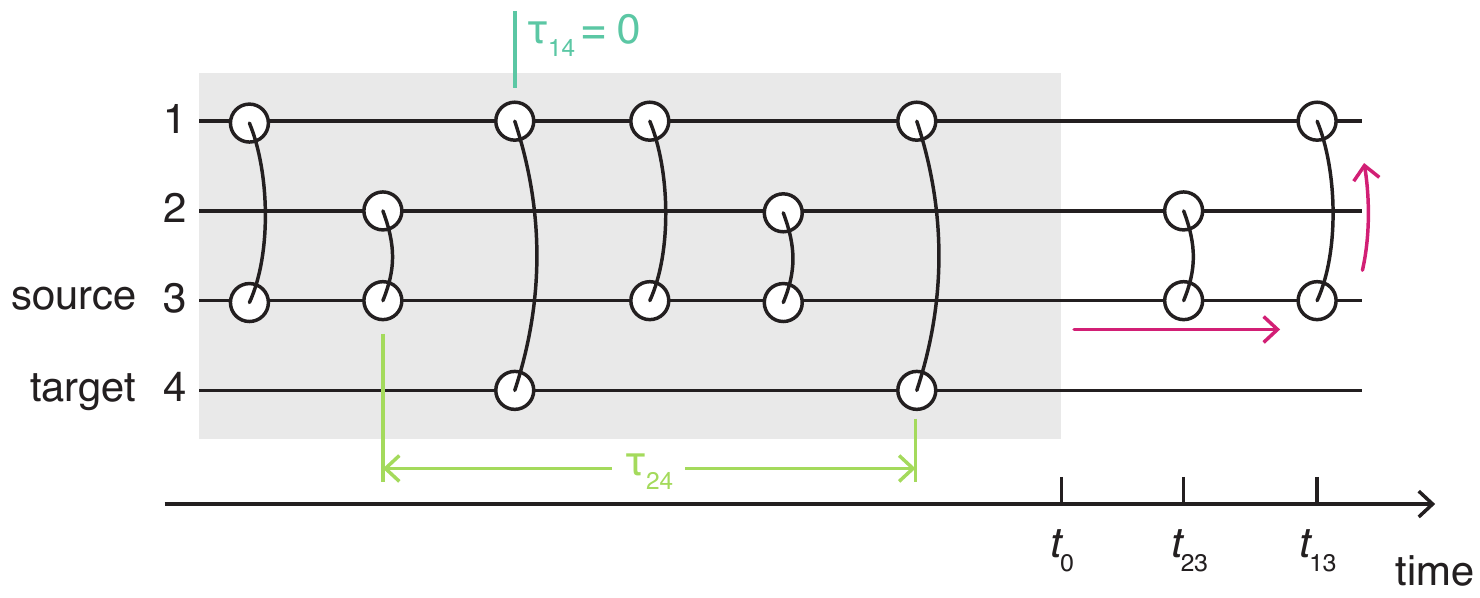} 
\caption{Schematic diagram of greedy navigation in temporal networks. 
Each horizontal line component corresponds to a node, horizontal axis represents time,
and vertically connected nodes represent contacts at that time~\cite{TemporalNetworkReview1,TemporalNetworkReview2,masudalambiotte}.
At time $t = t_0$, node $3$ (the source) 
wants to send a unit of package to node $4$ (the target).}
\label{fig:schematic}
\end{figure}

\subsection{Navigation}

Our model for temporal network navigation is inspired by our work on network navigation with spatial information~\cite{SHLee2012}. The idea for our \textit{greedy navigators} is that the agents know the direction of their target, and then takes steps as close  as possible (angularly) to its direction at every step.
In that case of spatial (and static) networks, therefore, the available information is embedded in the location of the nodes.
In the case of temporal networks, illustrated in Fig.~\ref{fig:schematic}, we consider the entire past or history as available information, and, for simplicity, assume that it is accessible to every node. 
During the run, we keep updating the information of the shortest path from a certain node to another.
For instance, in Fig.~\ref{fig:schematic}, node $3$ at $t = t_0$ estimates, based on its experience, that it takes one step from node $1$ to 
node $4$ and three steps (via node $3$ and $1$) from node $2$ to node $4$ in terms of the shortest hopping distance (the number of vertical jumps in the diagram), 
while $\tau_{1 \to 4}$ from node $1$ to node $4$
and $\tau_{2 \to 4}$ from node $1$ to node $4$ in terms of the shortest time duration (the horizontal time duration in such a diagram, assuming that
all of the interactions are instantaneous),
based on the history up to $t = t_0$.

The way our agents exploit the information---i.e.\ the way they are greedy---is realized by the following rule: if a walker at node $3$ tries to move to node $4$ at time $t = t_0$, as the node ``closest'' (the closest node is chosen uniformly at random in case of ties) to
the target (node $4$) is node $1$, node $3$ indefinitely waits for an interaction with node $1$ ($t = t_{13}$), even if the interaction between node $3$ and
node $2$ happens (at $t = t_{23}$) prior to the interaction between node $3$ and node $1$. Once the walker reached from node $3$ to node $1$, node $1$ will wait for
the direct interaction with node $4$ (one step or $\tau = 0$ to the target) and finalize the active navigation.

Our strategy assumes that historical interaction patterns 
will happen repeatedly or periodically and does no better job than random hopping in the absence of such temporal patterns (e.g., the Markovian Poisson process). On the other hand, in many empirical temporal networks
such periodical patterns exist due to the circadian rhythm and plays a crucial role in human interactions~\cite{HHJo2012}.
Note that we can define such a greedy navigation strategy based either on the hopping distance (distance-based temporal greedy navigation) or on the time (time-based temporal greedy navigation), as the node with the shortest distance to the target and the node with the shortest time to the target can be different. We will use both strategies as specified below and, when specificity is needed, denote the hop-based one by GNH and the time based one by GNT.

\subsection{Data}
\label{sec:data}

\begin{table}
\caption{Description of the basic properties of the data sets: the number of nodes $N$, the number of contacts $C$, the duration $T$, the number of samples $n$, and the time resolution $\delta t$. \label{tab:props}}
\begin{tabular}{lllllll}
Data set & $N$ & $C$ & $T$ & $n$ & $\delta t$ & Ref. \\ \hline
Gallery & $159.01$ & $6027.71$ & $1$ day & $69$ & $20$ seconds & \cite{gallery} \\
Primary School & $242$ & $125773$ & $1$ day & $1$ & $20$ seconds & \cite{school} \\
High School 1 & $126$ & $28561$ & $4$ days & $1$ & $20$ seconds & \cite{hschool} \\
High School 2 & $180$ & $45047$ & $7$ days & $1$ & $20$ seconds & \cite{hschool} \\
Conference & $113$ & $20818$ & $2.5$ days & $1$ & $20$ seconds & \cite{conference} \\
Hospital & $75$ & $32424$ & $4$ days & $1$ & $20$ seconds & \cite{Vanhems2013} \\
\end{tabular}
\end{table}

In this section, we will discuss the empirical data sets we use. All our data sets come from the SocioPatterns project \url{sociopatterns.org} where human proximity interactions are tracked by radio-frequency identification devices
in various types of locations such as a hospital, a primary school, a high school an art gallery, and conference. Some of the data sets come from repeated experiments over $n$ days. In this case, we run our analyses on the separate days and average the results. A summary of the basic properties of these data sets and references to the original studies can be found in Table~\ref{tab:props}.

\section{Results}
\label{sec:results}

We will start with an example---navigation on the \textit{Hospital} data set---then continue to aggregate properties of all data sets.

\subsection{Different strategies applied to the Hospital data}
\label{sec:hospital_data}

\begin{figure}
\includegraphics[width=0.9\columnwidth]{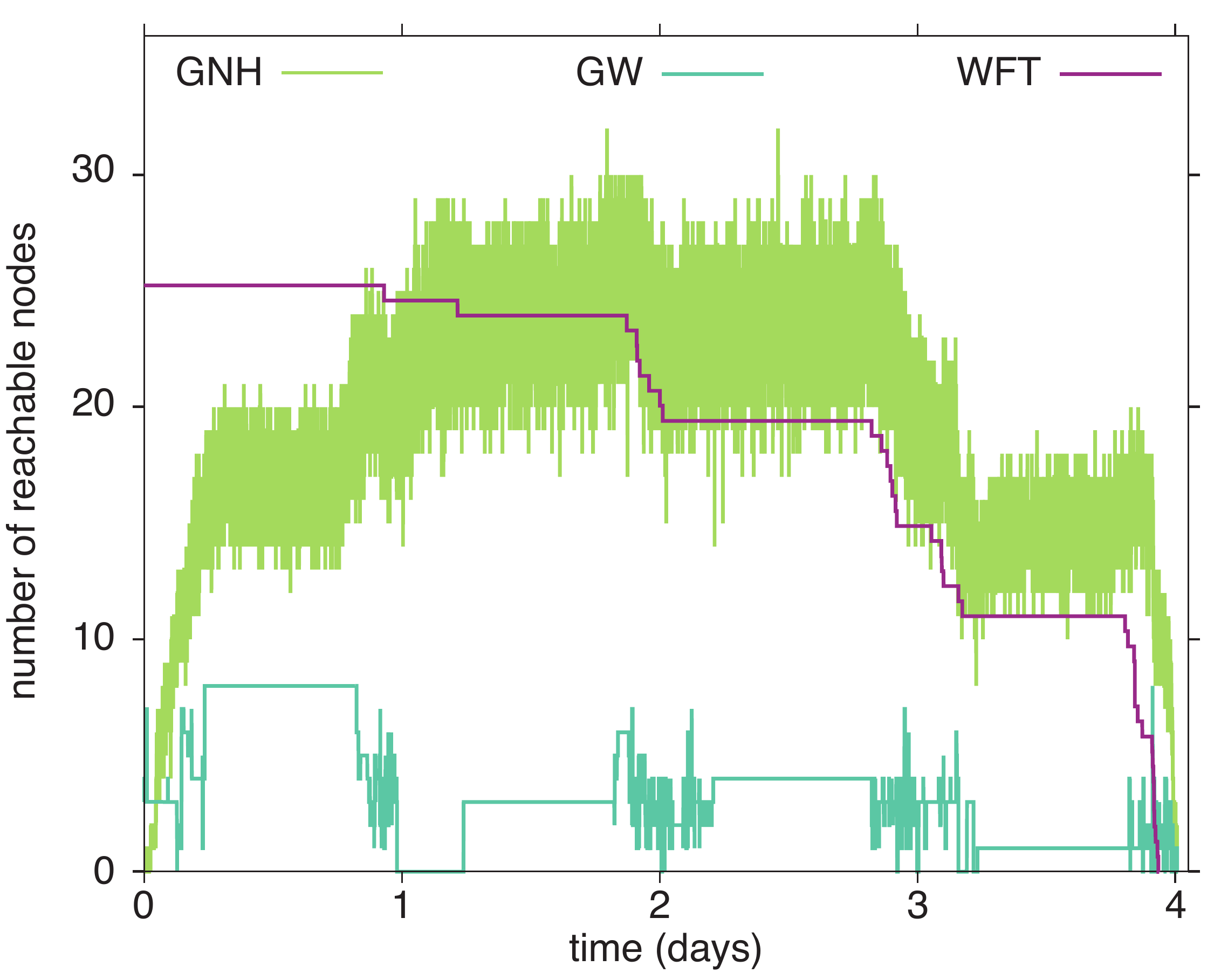}
\caption{A plot of the reachability versus average time to all of the other nodes from a certain node (denoted as ``node $0$'') of the hospital ward dynamic contact network at each time. For all of the cases, the time unit is day.}
\label{fig:reachability_hospital}
\end{figure}

\begin{figure}
\includegraphics[width=0.9\columnwidth]{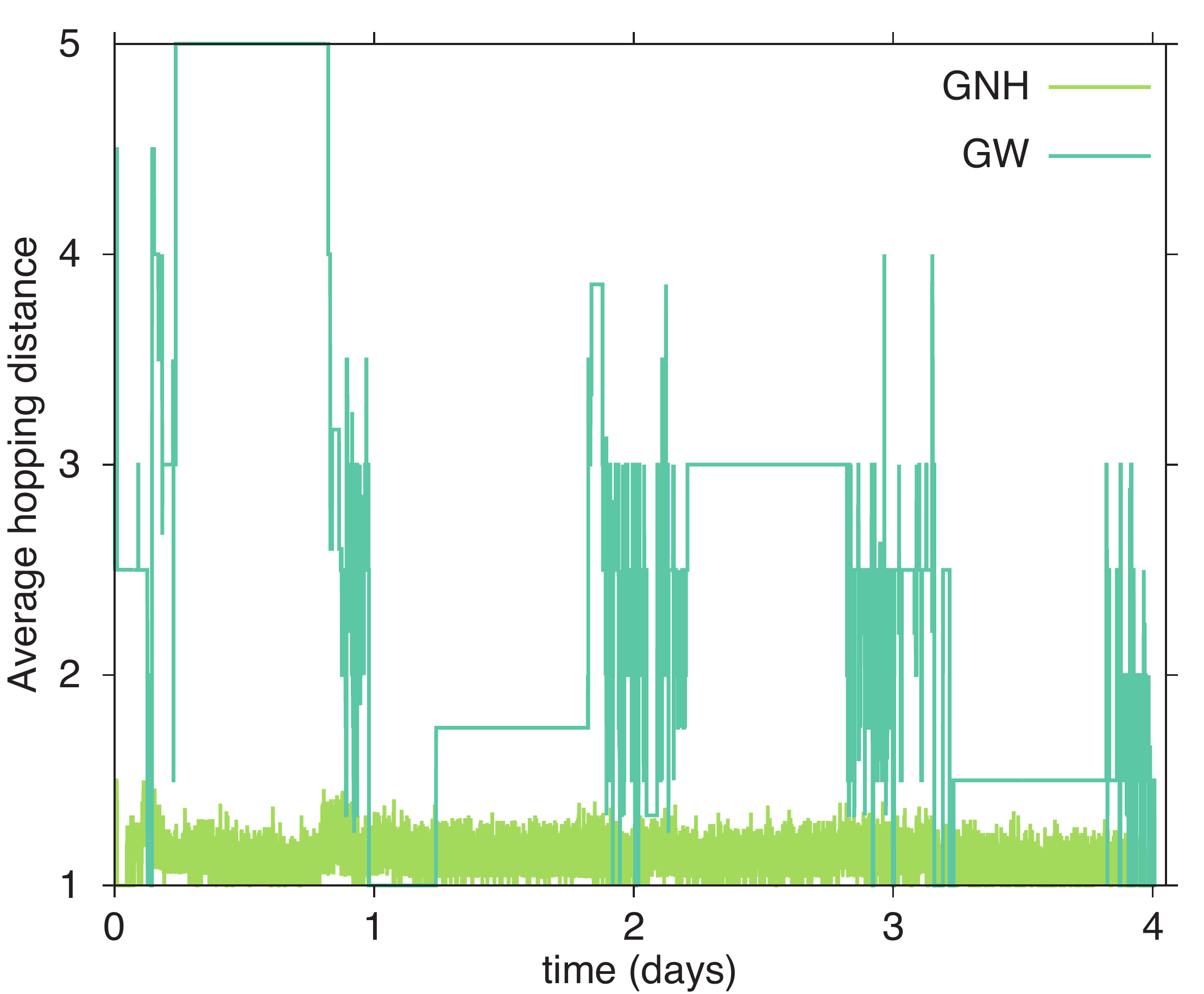}
\caption{A plot corresponding to Ref.~\ref{fig:reachability_hospital} but for the average distance in the navigation.
}
\label{fig:distance_hospital}
\end{figure}

\begin{figure}
\includegraphics[width=0.9\columnwidth]{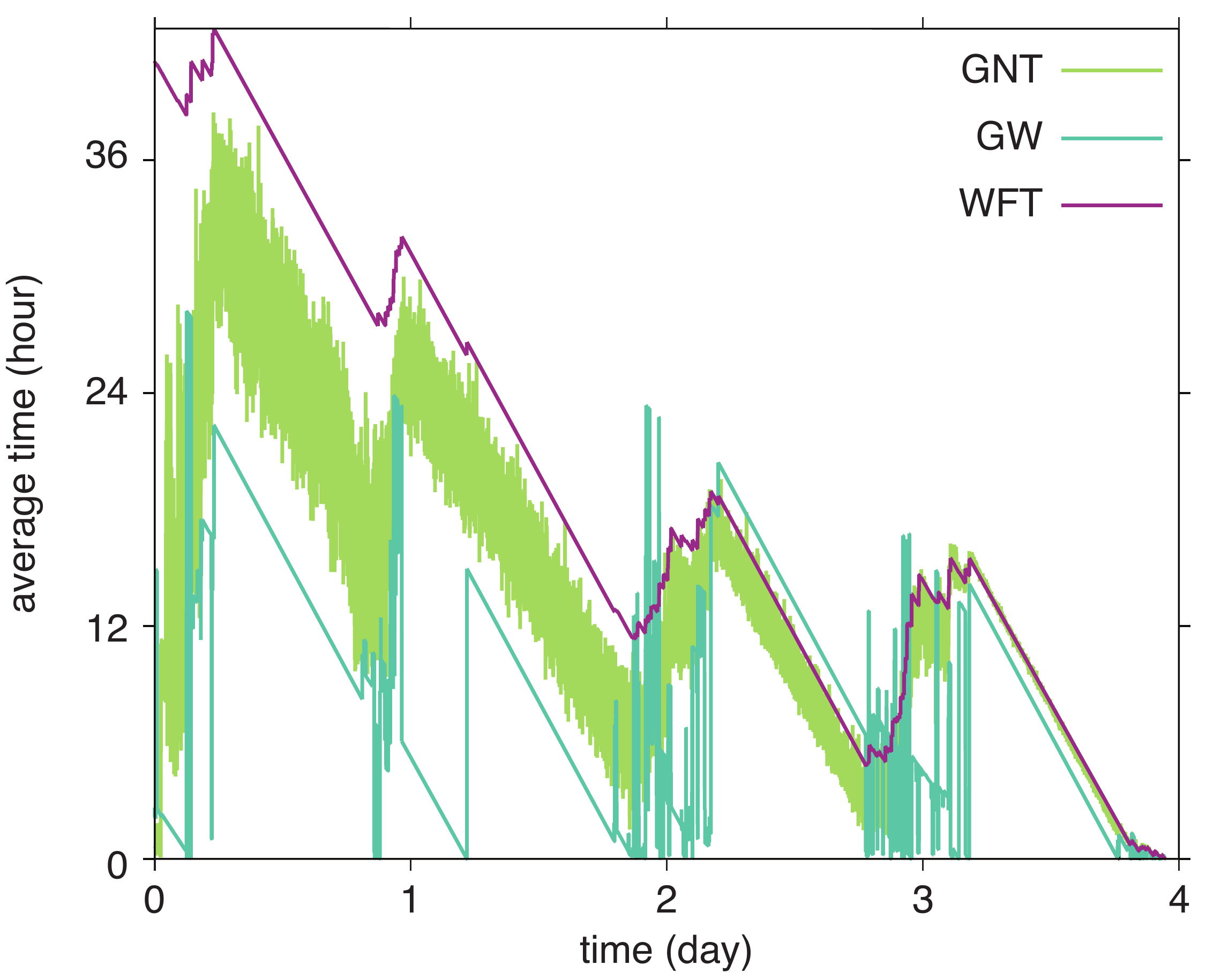}
\caption{A plot corresponding to Ref.~\ref{fig:reachability_hospital} but for the average time to the target.}
\label{fig:time_hospital}
\end{figure}

Figure~\ref{fig:reachability_hospital} shows the reachability---the fraction of reachable nodes in the future that can be connected by time-respecting paths~\cite{TemporalNetworkReview1}---for a given starting nodes as a function of the time of the beginning of the navigation for the Hospital network~\cite{Vanhems2013}. We see that the greedy navigators (the hop-based version) has the largest reachability for intermediate values of the time. The time based greedy navigation is very similar and omitted for clarity. The waiting-for-target strategy is monotonically decreasing, which is trivial as those agents do nothing to improve their chance of reaching the target. Greedy navigators are always more efficient than the greedy walkers, whereas WFT is the best strategy for early times. At the early times greedy navigators have not assembled enough information to outperform greedy walkers. Thus, a yet more efficient strategy might thus be to wait for the target in the beginning of the time period, then switch to greedy navigation after a while. At the end of the data, the reachability goes down for all strategies simply because the number of time-respecting paths goes to zero.

The results for the average hopping distances (the number of hops to reach the target, given that the target is reached) are plotted in Fig.~\ref{fig:distance_hospital}. In this case the WFT strategy is trivially one so we omit it from the figure. The greedy navigators are always just a little over one, indicating that the mostly just wait for the target. The greedy walkers have a complex pattern that comes from that their character of always moving in combination that the movement can be almost deterministic for walks with few ties.

Next, we turn to the time-based greedy navigators. We plot the time to reach the target (given that the target is reached) for the Hospital data set in Fig.~\ref{fig:time_hospital}. This is the corresponding quantity to hop-distance of Fig.~\ref{fig:distance_hospital} for this time-based picture. This figure show the typical saw-tooth pattern of latency (the minimal possible time to reach from one node to another). The time goes down linearly until a key contact is passed, then it jumps discontinuously upward by the time to the next contact on a path to the target.

\begin{figure*}
\includegraphics[width = 0.5\textwidth]{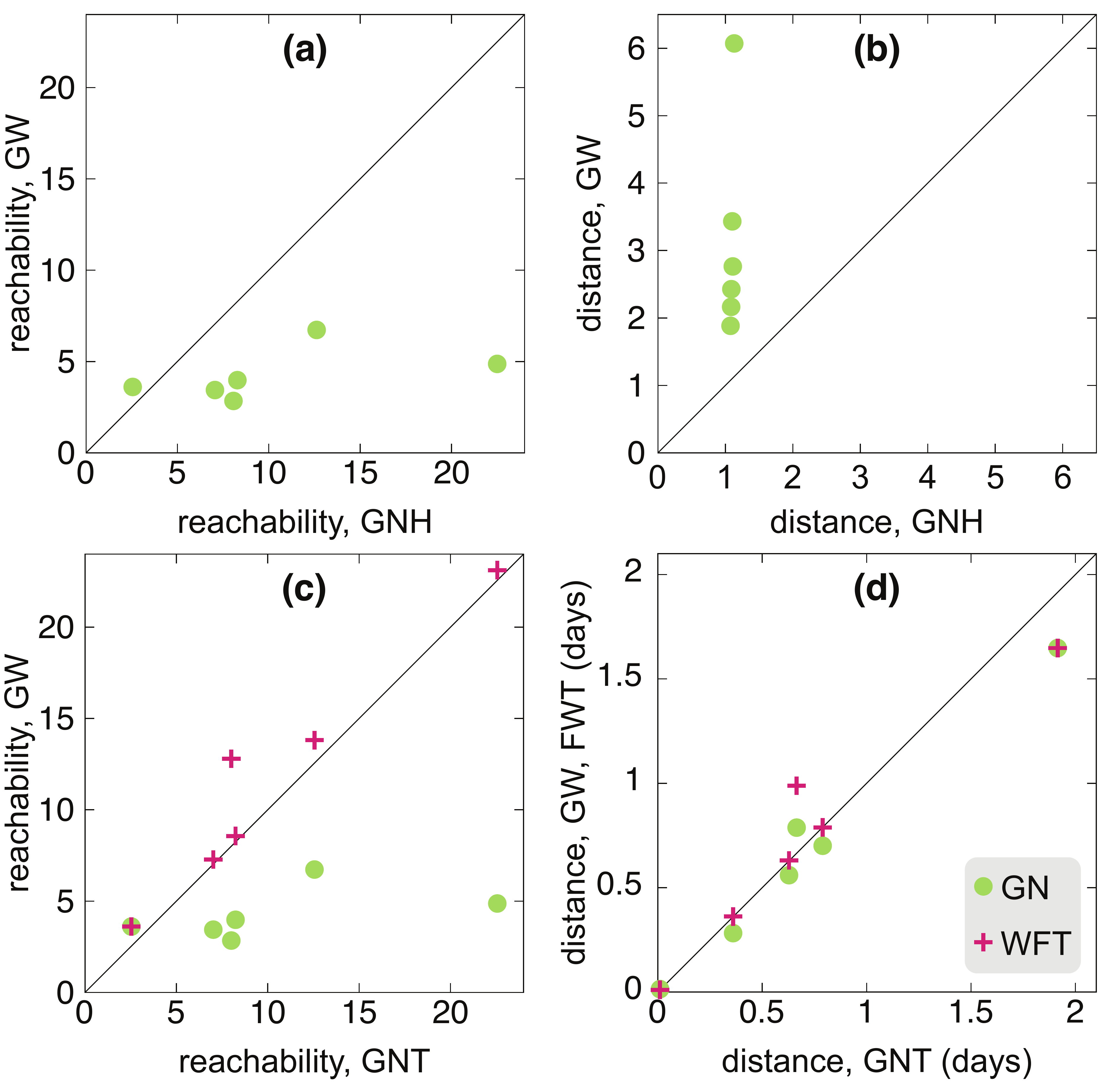}
\caption{Average values of descriptive quantities of the navigation for our six data sets. Panel (a) plots the average reachability of hop-based greedy walkers as a function of the average reachability of greedy navigators. The diagonal line shows where the two quantities equal each other. Panel (b) is the corresponding figure for the hopping-count distance of the navigations. Panel (c) shows the average reachability  for time-based navigations. In this panel, we also include values for the WFT strategy (in addition to greedy walkers). Panel (d) shows the time to reach the target for the time-based navigators. This is the corresponding measure to the distance of panel (b).
}
\label{fig:averages}
\end{figure*}

\begin{figure*}
\includegraphics[width = 0.6\textwidth]{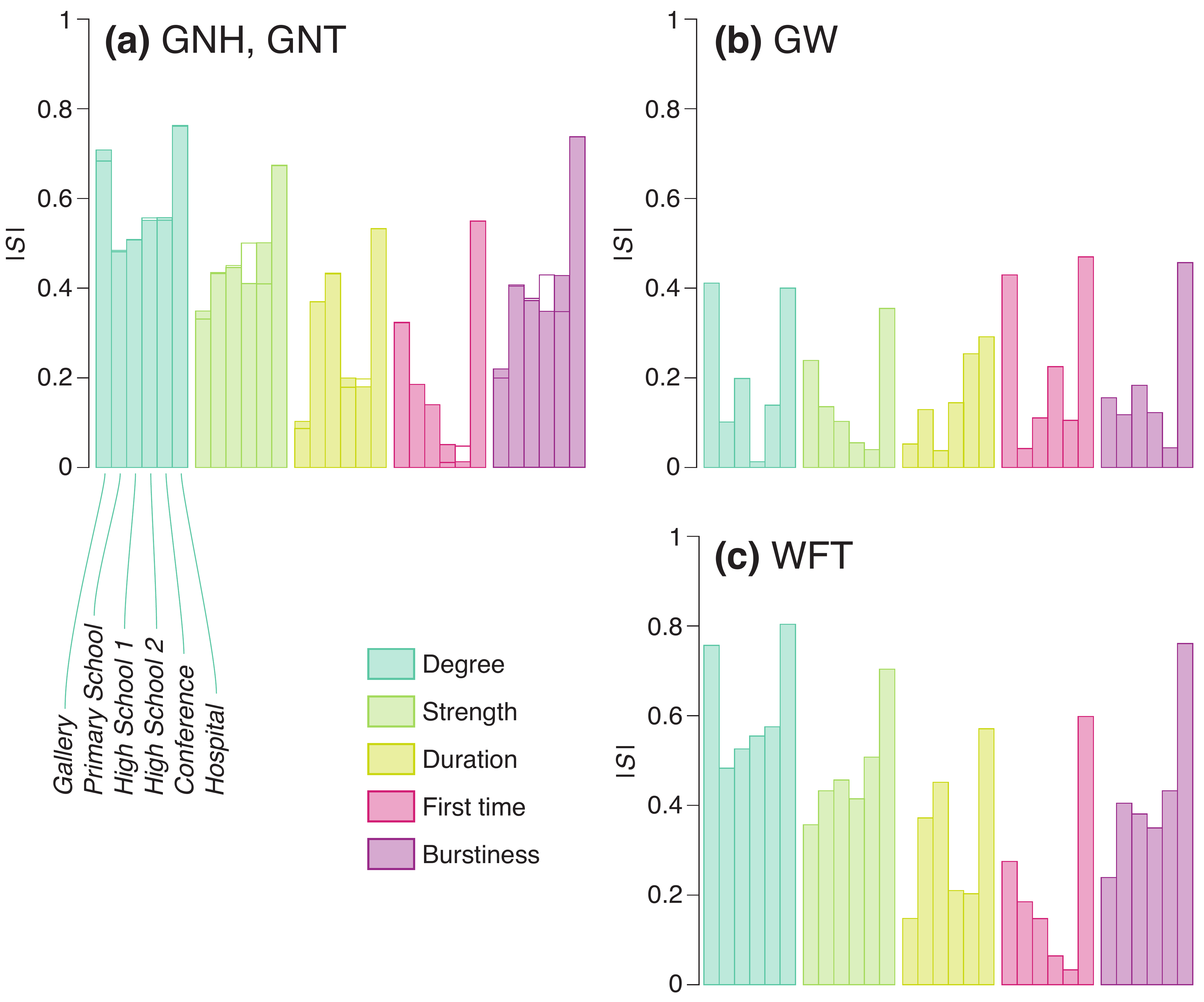}
\caption{The Spearman correlation between quantities describing the positions of nodes in the network---degree, strength, duration, age and burstiness---and the reachability. The outlined bars corresponds to GNT. Panel (a) shows results for greedy navigators; (b) the corresponding for greedy walkers and (c) for wait-for-target walkers.
}
\label{fig:corr_hop_reachability}
\end{figure*}

\begin{figure}
\includegraphics[width = \columnwidth]{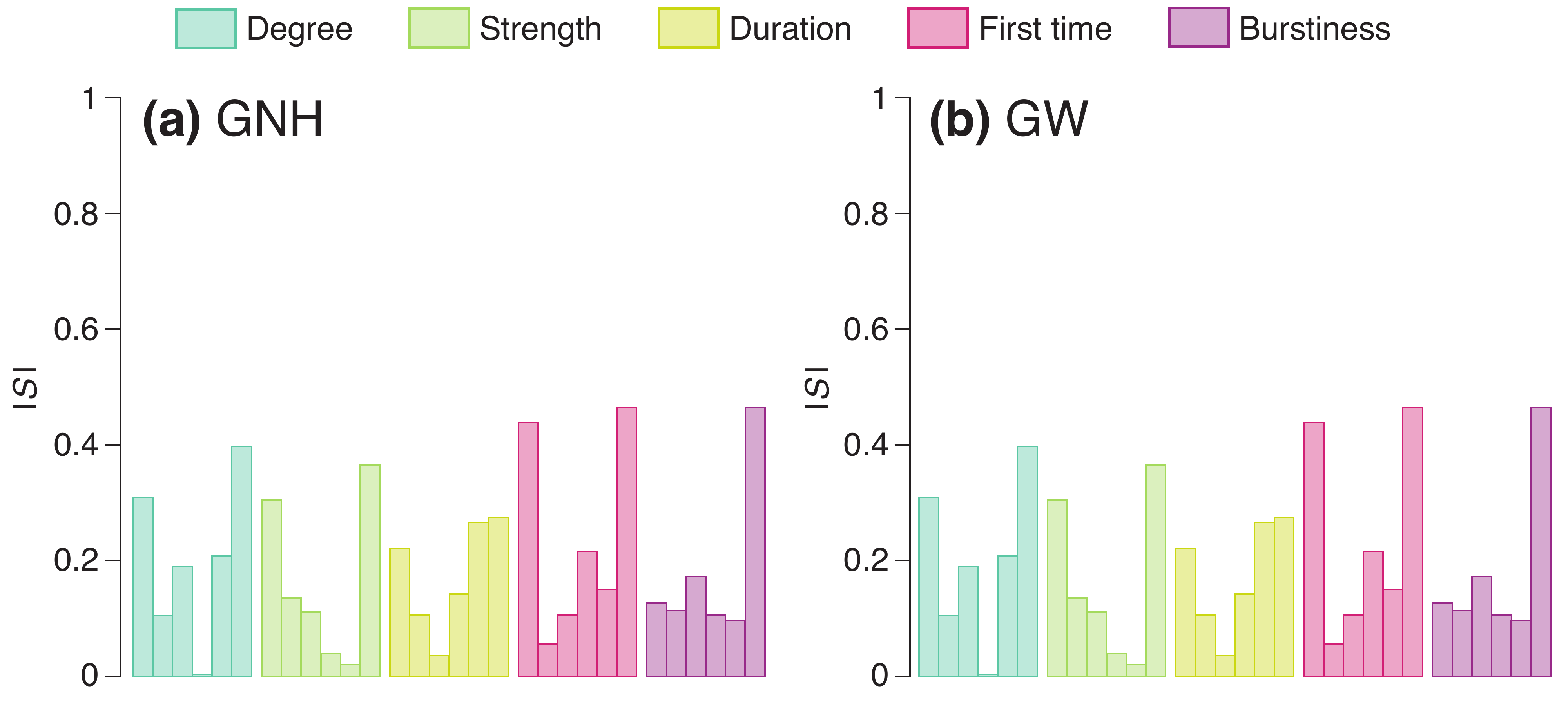}
\caption{Plots corresponding to Fig.~\protect{\ref{fig:corr_hop_reachability}} but for the average hopping distance.} This quantity is trivially one for WFT and thus not shown.
\label{fig:corr_hop_distance}
\end{figure}

\begin{figure*}
\includegraphics[width = 0.6\textwidth]{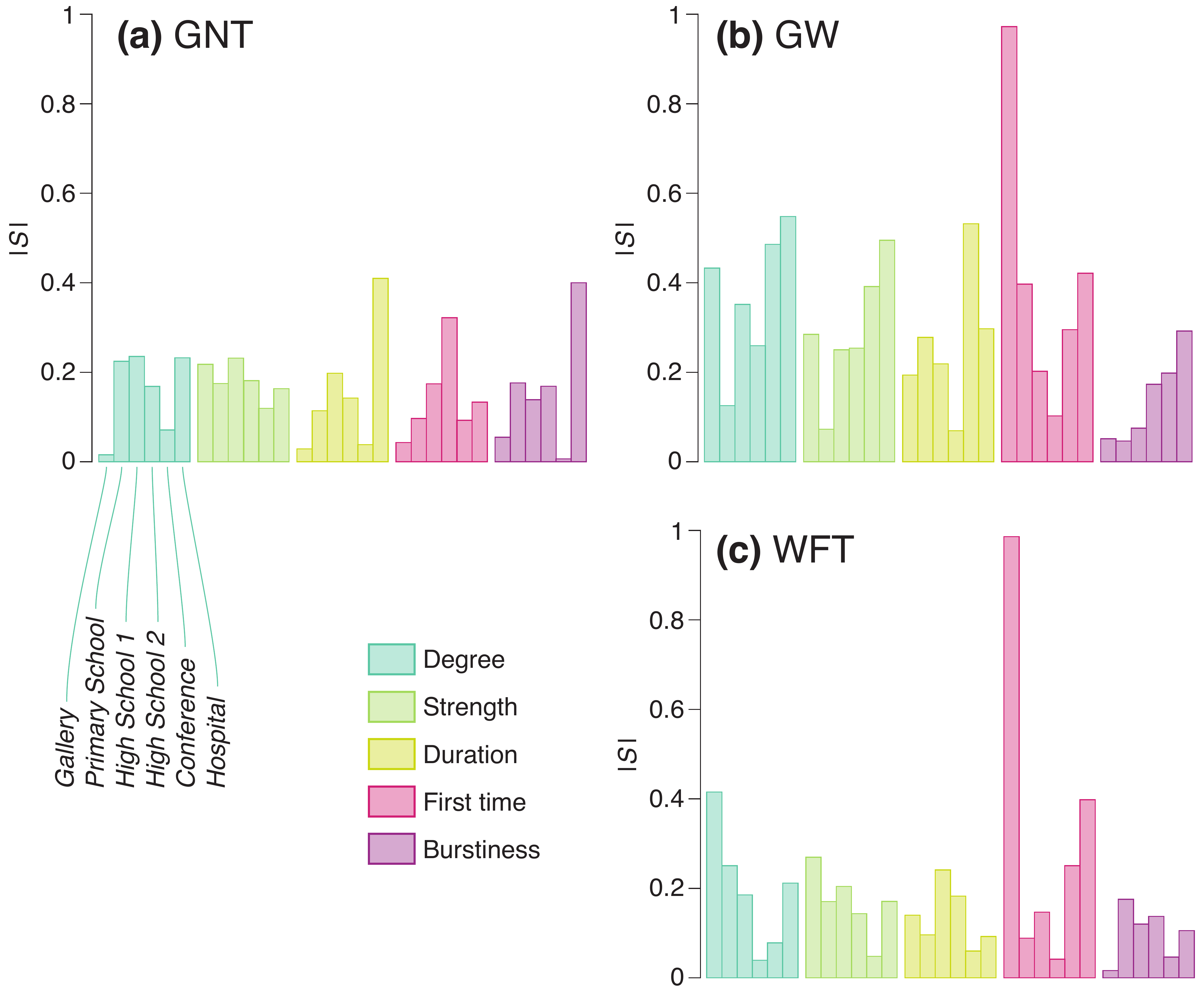}
\caption{Plots corresponding to Fig.~\protect{\ref{fig:corr_hop_reachability}} but for the average time to the target.
}
\label{fig:corr_time_distance}
\end{figure*}

\subsection{Average navigation performance}

In Figs.~\ref{fig:reachability_hospital}, \ref{fig:distance_hospital} and \ref{fig:time_hospital}, we see the navigation performance from one node in one networks. Next, we plot the average values of these quantities over all nodes and all data sets. See Fig.~\ref{fig:averages}. We plot the three quantities of the greedy walks and wait-for-target strategy as a function of the corresponding values for greedy navigation. We see that for all the data sets, except one (Gallery), the greedy navigation outperforms greedy walks. The exception has a peculiar interaction structure. Since it records visitors coming and going to an art gallery, the people present early in the data are not present at the end. This means that relying on the past reachability information is not only useless, it is misleading---if the best path to the target went via node $i$ before, then $i$, by a large chance left the gallery already. We can see that the hop-based and time-based strategies (of panel (c)) are very similar.

In Fig.~\ref{fig:distance_hospital}(b), we can see the distance (number of contact followed during the navigation) for the two non-trivial strategies GN and GW (WFT is constantly one). Although it is hard to prove that there is no temporal network data set where the distance is longer for GN than GW, it sounds very unlikely that an empirical (relatively well-behaved) data set would have that feature. Interestingly, the GN values are closer to one than two. This means that greedy navigators usually wait for the target, and only occasionally exploits their information to improve the navigation. This makes the increase in performance as seen in Fig.~\ref{fig:distance_hospital}(a) quite remarkable.

In Fig.~\ref{fig:distance_hospital}(c), we plot the time to the target, given that the target is reached. This quantity is quite similar for all three strategies (which we could also see in Fig.~\ref{fig:time_hospital}). One explanation is that if the agents find a path, they typically find the same one.

\subsection{Structural explanations of navigability}
\label{sec:correlation}

Navigability, as reflected in our three measured quantities, can change abruptly. The variability among nodes is also very large. In this section, we explore if this variability can be explained by temporal-network measures quantifying the position of nodes. We try a mix of static and temporal quantities: The number of other nodes a node is in contact with (\textit{degree}), the number of contacts a node participates in (\textit{strength}), the time between the first and last contact a node participates in (\textit{duration}), the time of the first contact of a node (\textit{first time}). Finally, we measure the \textit{burstiness} as defined in Ref.~\cite{KIGoh2008}:
\begin{equation}
B \equiv \frac{\sigma_\tau - m_\tau}{\sigma_\tau + m_\tau} \,,
\label{eq:GB_burtiness}
\end{equation}
where $m_\tau$ and $\sigma_\tau$ are the mean and the standard deviation of $P(\tau)$ (the distribution of times $\tau$ between contacts of a node). We use the absolute value $|S|$ of the Spearman rank correlation  since several of the quantities have heavy-tailed distributions.

In Fig.~\ref{fig:corr_hop_reachability}, we plot the $S$ for the reachability values and the five positional descriptors. We see that the values are rather large, so these simple structures can explain the behavior of the navigators to a fairly large extent. Then we notice that degree and burstiness give slightly larger values of the correlation compared to other structural measures; however, the variation is not extremely large---other measures are also correlated. This can to some extent be explained by them being correlated to each other---if e.g.\ the degree is large, then so is probably also the strength. It is interesting that degree and burstiness are better predictors since they measure two very different aspects. The first time shows a bit lower values of $|S|$. The navigation problem is thus dependent on both temporal and topological structures. This is similar other dynamics on the network~\cite{masuda_holme_rev}. We also note that GNH and GNT give very similar values. In practice there seems to be no point in separating them.

Greedy walks (Fig.~\ref{fig:corr_hop_reachability}(b)) are less correlated with the structural measures, whereas WFT agents show a similar behavior to GN. Of the data sets, Hospital has the highest correlation values for all structural measures. The data set with the weakest correlations vary with the structural measures. For the first time statistics, Gallery shows high correlations. As mentioned, this could be understood from the time stretched nature of this datasets---the first time correlates with the presence of a node in the end and beginning of the data. 

In Fig.~\ref{fig:corr_hop_distance}, we show the correlations between the hopping distance and the structural quantities. These are somewhat weaker than the reachability (ranging between 0 and 0.4 as opposed to 0 to 0.8 for reachability. The correlations for WFT are very similar to those of GN; see Fig.~\ref{fig:corr_hop_distance}(c). Also in this case, the Gallery and Hospital data sets are the ones with highest correlations.

In our final analysis plot, Fig.~\ref{fig:corr_time_distance}, we investigate correlations with the time to reach a reachable target. Over all, this case gives low correlations. This could be understood since many of the actual paths leading to the target are rather few and the waiting time between the contacts depend on many outer, effectively random factors. In this case, the Gallery data shows spectacularly large values of the correlations for GW and WFT with the first times. Once again, this can be understood from the time-stretched nature of the network. There are relatively long paths from sources in the early times of the data to targets in the end. Since greedy navigation bases its paths on (in this case, misinformed) data, it shows weak correlations.

\section{Discussion and Conclusions}
\label{sec:discussion}

In this paper, we have introduced the problem of navigation in temporal networks as the problem to decide whether or not to follow a contact given what is known of past interactions. We have contrasted a temporal-network navigation strategy---greedy navigation---that assumes the past predicts the future, with two more simplistic strategies.

Greedy navigation is a very simple strategy, just taking steps that would have worked well in the past, disregarding any trends in the activity of the nodes, etc. Still one would expect real agents in a temporal network context to have a similarly simple intuitive approach (rather than some strategy with heavy computational overhead). On the other hand, it is hard to think of real processes that are very well described by our scenario. A subway passenger without a map, could be a mental scenario, though thoroughly unlikely. Some distributed computing problems could probably also be candidate applications---perhaps opportunistic networking where data is transferred through devices in close proximity~\cite{huang2008survey}.

Even though the information accessible to the greedy navigators is quite elementary (the list of temporal distance between the target and all of the other nodes), we have demonstrated that they can actually augment the path-finding process by exploiting temporal correlations. There is thus information encoded in the past contacts that can be exploited in future navigation. There is not one structural quantity that can explain the behavior of our navigators---some structures (degree and burstiness) do correlate more strongly with the quantities describing the navigation; some data sets have consistently stronger correlations than other. Finally, it is easier to explain the reachability in terms of the position of nodes, than it is to describe the number of hops, or the time to reach the target (if it is reached).

We believe this is the beginning of an interesting research direction of temporal network research. One can extend our research by finding more efficient, simple navigation strategies. It could be interesting to vary the accessible information, or to tune the structures in a systematic way in a model-based study.

\bibliographystyle{abbrv}
\bibliography{navi}

\end{document}